# When Database Systems Meet the Grid

María A. Nieto-Santisteban
Alexander S. Szalay
Aniruddha R. Thakar
William J. O'Mullane

Johns Hopkins University

Jim Gray

Microsoft Research

James Annis

Experimental Astrophysics, Fermilab

# When Database Systems Meet the Grid

María A. Nieto-Santisteban[1], Jim Gray[2], Alexander S. Szalay[1], James Annis[3], Aniruddha R. Thakar[1], and William J. O'Mullane[1]
1. Johns Hopkins University, Baltimore, MD, USA
2. Microsoft Research, San Francisco, CA, USA
3. Experimental Astrophysics, Fermilab, Batavia, IL, USA
nieto, szalay, thakar, womullan@pha.jhu.edu,  gray@microsoft.com, annis@fnal.gov

**Abstract**

We illustrate the benefits of combining database systems and Grid technologies for data-intensive applications. Using a cluster of SQL servers, we reimplemented an existing Grid application that finds galaxy clusters in a large astronomical database. The SQL implementation runs an order of magnitude faster than the earlier Tcl-C-file-based implementation. We discuss why and how Grid applications can take advantage of database systems.

## 1. Introduction

Science faces a data avalanche. Breakthroughs in instruments, detector and computer technologies are creating multi-Terabyte data archives in many disciplines. Analysis of all this information requires resources that no single institution can afford to provide.  In response to this demand, Grid computing has emerged as an important research area, differentiated from clusters and distributed computing. Many definitions of the Grid and Grid systems have been given [17]. In the context of this paper, we think of *the Grid as the infrastructure and set of protocols that enable the integrated, collaborative use of high-end computer, networks, databases, and scientific instruments owned and managed by multiple organizations, referred to virtual organizations* [18][27].

The need to integrate databases and database technology into the Grid was already recognized, in order to support science and business database applications as well as to manage metadata, provenance data, resource inventories, etc. [16].  Significant effort has gone into defining requirements, protocols and implementing middleware  to access databases in Grid environments [19][20][21][22][23]. Although database management systems (DBMS) have been introduced as useful tools to manage metadata, data, resources, workflows, etc [24][25][26], the presence of databases is minimal in *science* applications running on the Grid.  Today the typical data-intensive science Grid application still uses flat files to process and store the data and cannot benefit from the power that database systems offer.

To evaluate the benefit of combining database and Grid technologies, this paper compares an existing file-based Grid application, MaxBCG [6], with an equivalent SQL implementation.  This paper describes the MaxBCG algorithm and its relationship to the Sloan Digital Sky Survey (SDSS) and the Virtual Observatory (VO) project. Next, we describe in detail the file-based and database implementations, and compare their performance on various computer systems. Finally, we discuss how the SQL implementation could be run efficiently on a Grid system.  We conclude by speculating why database systems are not being used on the Grid to facilitate data analysis.

## 2. Finding Galaxy Clusters for SDSS

Some Astronomy knowledge is needed to understand the algorithm's computational requirements [28]. Galaxies may be categorized by brightness, color, and redshift. Brightness is measured in specific wavelength intervals of light using standard filters. Color is the difference in brightness through two different filters. Due to the Hubble expansion of the Universe, the Doppler redshift of light from a galaxy is a surrogate for its distance from Earth.

Galaxy clusters are collections of galaxies confined by gravity to a compact region of the universe. Galaxy clusters are useful laboratories for studying the physics of the Universe. Astronomers are developing interesting new ways to find them systematically. The brightest galaxy in a cluster (BCG) is typically the most massive and so tends to be near the cluster center.

The Maximum-likelihood Brightest Cluster Galaxy algorithm [1], MaxBCG, finds galaxy clusters. It has been used to search the Sloan Digital Sky Survey (SDSS) catalog for Cluster candidates [2]. MaxBCG was originally implemented as Tcl scripts orchestrating the SDSS Astrotools package [3] and ran on the Terabyte Analysis Machine (TAM), a 5-node Condor cluster specifically tuned to solve this type of problem [4][5]. The same application code was integrated with the Chimera Virtual Data System created by the Grid Physics Network (GriPhyN) project to test Grid technologies [6]. As is common in astronomical file-based Grid applications, the TAM and Chimera implementations use hundreds of thousands of files fetched from the SDSS Data Archive Server (DAS) to the computing nodes.

SkyServer is the Web portal to the SDSS Catalog Archive Server (CAS) – the relational database system hosting the SDSS catalog data. All the data required to run MaxBCG is available in the SkyServer database. SDSS is part of the Virtual Observatory also known as the World Wide Telescope. The Virtual Observatory is being implemented in many countries [7]. It is developing portals, protocols, and standards that federate and unify many of the world's astronomy archives into a giant database containing all astronomy literature, images, raw data, derived datasets, and simulation data integrated as a single intelligent facility [8].

The World-Wide Telescope is a prototypical data Grid application supporting a community of scholars cooperating to build and analyze a data Grid that integrates all astronomy data and literature. The MaxBCG search for clusters of galaxies is typical of the tasks astronomers will want to perform on this data Grid.

### 2.1 The Algorithm

The MaxBCG algorithm solves the specific astronomical problem of locating clusters of galaxies in a catalog of astronomical objects. It searches for galaxy clusters over a wide range of redshifts and masses. The search relies on the fact that the brightest cluster galaxies (BCG) in most clusters have remarkably similar luminosities and colors [9]. The MaxBCG algorithm works on a 5-dimensional space and calculates the cluster likelihood of each galaxy. The 5-space is defined by two **spatial** dimensions, Right Ascension, ra, and Declination, dec; two **color** dimensions, g-r and r-i; and one **brightness** dimension, i. The algorithm includes six steps:

**Get galaxy list** extracts the five-dimensions of interest from the catalog.
**Filter** calculates the unweighted BCG likelihood for each galaxy (unweighted by galaxy count) and discards unlikely galaxies.
**Check neighbors** weights the BCG likelihood with the number of neighbors.
**Pick most likely** for each galaxy, determines whether it is the most likely galaxy in the neighborhood to be the center of the cluster.
**Discard compromised results** removes suspicious results and stores the final cluster catalog.
**Retrieve the members of the clusters** retrieves the galaxies that the MaxBCG algorithm determined are part of the cluster.

### 2.2 The TAM Implementation

The MaxBCG algorithm was implemented as Tcl scripts driving Astrotools, which is an SDSS software package comprised of Tcl and C routines layered over a set of public domain software packages [3]. The CPU intensive computations are done by Astrotools using external calls to C routines to handle vector math operations. The algorithm ran on the TAM Beowulf cluster [4].

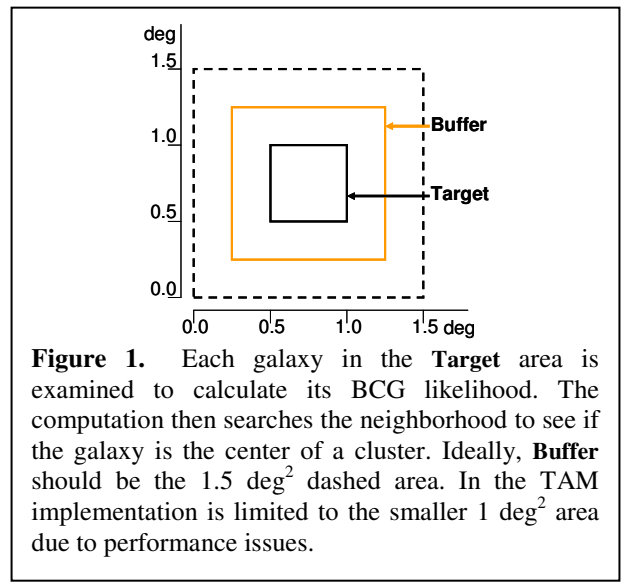

**Figure 1.** Each galaxy in the **Target** area is examined to calculate its BCG likelihood. The computation then searches the neighborhood to see if the galaxy is the center of a cluster. Ideally, **Buffer** should be the 1.5 deg$^2$ dashed area. In the TAM implementation is limited to the smaller 1 deg$^2$ area due to performance issues.

The TAM MaxBCG implementation takes advantage of the parallel nature of the problem by using a divide-and-conquer strategy which breaks the sky in 0.25 deg$^2$ fields. Each field is processed as an independent task. Each of these tasks require two files: a 0.5 x 0.5 deg$^2$ **Target** file that contains galaxies that will be evaluated and a 1 x 1 deg$^2$ **Buffer** file with the neighboring galaxies needed to test for the presence of a galaxy cluster. Ideally the **Buffer** file would cover 1.5 x 1.5 deg$^2$ = 2.25 deg$^2$ to find all neighbors within 0.5 deg of any galaxy in the **Target** area and estimate the likelihood that a galaxy is the brightest one in a cluster. But the time to search the larger **Buffer** file would have been unacceptable because the TAM nodes did not have enough RAM storage to hold the larger files: the compromise was to limit the buffer to cover only to 1 x 1 deg$^2$ areas [Figure1].

A **Target** field of 0.25 deg$^2$ contains approximately 3.5 x 10$^3$ galaxies. Initially, every galaxy in the catalog is a possible BCG. The observed brightness and color of each candidate is compared with entries in **a k-correction** table, which contains the expected brightness and color of a BCG at 100 possible redshifts. This comparison yields a (perhaps null) set of plausible redshifts for each candidate BCG. If, at any redshift, a galaxy has even a remote chance of being the right color and brightness to be a BCG, it is passed to the next stage.

Given a candidate galaxy, the next stage uses the **Buffer** file to compute the number of neighbor galaxies at every redshift. This *every redshift* search is required because the color window, the magnitude window, and the search radius all change with redshift. The BCG likelihood is computed at each redshift. The maximum likelihood, over the entire range of redshifts for the object with at least one neighbor, is recorded in the BCG **Candidates** file, **C**. About 3% of the galaxies are candidates to be a BCG.

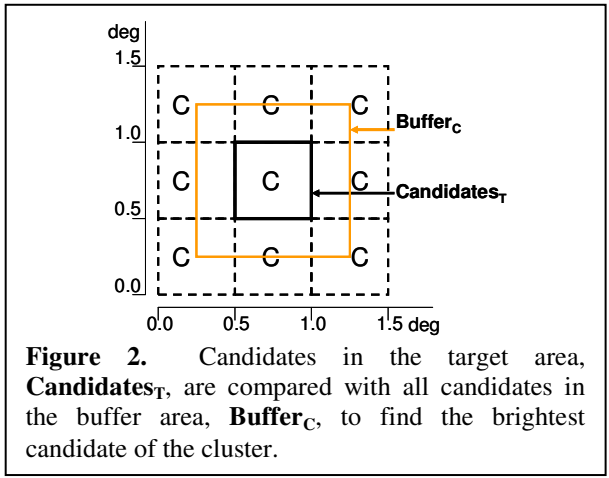

**Figure 2.** Candidates in the target area, **Candidates$_T$**, are compared with all candidates in the buffer area, **Buffer$_C$**, to find the brightest candidate of the cluster.

In order to determine whether a candidate galaxy is a BCG, rather than just a member of the cluster, the algorithm compares it with the neighboring **candidates** which are compiled into the **Buffer$_C$** file [Figure 2]. Ideally, each candidate should be compared with all candidates within 0.5 deg as this corresponds to a reasonable low redshift cutoff. However, as explained earlier [Figure 1], TAM is restricted to 1 x 1 deg$^2$ area to meet its computation time and storage budget, leaving only a 0.25 deg buffer surrounding the 0.5 x 0.5 deg$^2$. The algorithm finds approximately 4.5 clusters per target area (0.13% of the galaxies are BCGs).

The last step is to retrieve the galaxies in the cluster. A galaxy is considered to be part of the cluster if it is inside a radius of 1 Mpc (3.26 million light years, converted into degrees using the redshift) of the BCG and inside the *R200* radius containing 200 times the background mass density. The *R200* radius is derived from the cluster mass (number of galaxies) using a lookup table. In the TAM implementation these spherical neighborhood searches are reasonably expensive as each one searches the **Buffer** file.

Once the **Buffer** and **Target** files are loaded into RAM the algorithm is CPU-bound. The 600 MHz CPUs of the TAM could process a **Target** field of 0.25 deg$^2$ in about a thousand seconds. Processing the many target fields is embarrassingly parallel, so the time scales lineally with the number of target areas being processed. TAM is composed of 5 nodes, each one a dual-600-MHz PIII processor nodes each with 1 GB of RAM. The TAM cluster could process ten target fields in parallel.

### 2.3 SQL Server DBMS Implementation

We implemented the same MaxBCG algorithm using the SDSS CAS database [10]. This new implementation includes two main improvements. First, it uses a finer k-correction table with redshift steps of 0.001, instead of 0.01. Second, it uses a 0.5 deg buffer on the target field. Although these two improvements give better scientific results, would have increased the TAM processing time by a factor of about 25. The implementation is available from [29].

As described in Section 2.2, the TAM approach builds two files, **Target** and **Buffer**, for each 0.25 deg$^2$ target field. The SQL application processes much larger pieces of the sky all at once. We have been using a target area of 11 deg x 6 deg = 66 deg$^2$ inside a buffer area of 13 deg x 8 deg = 104 deg$^2$; but, in principle the target area could be much larger. Larger target areas give better performance because the relative buffer area (overhead) decreases [Figure 3]. Using a database and database indices allows this much large area because the database scans the areas using high-speed sequential access and spatial indices rather than keeping all the data in the RAM.

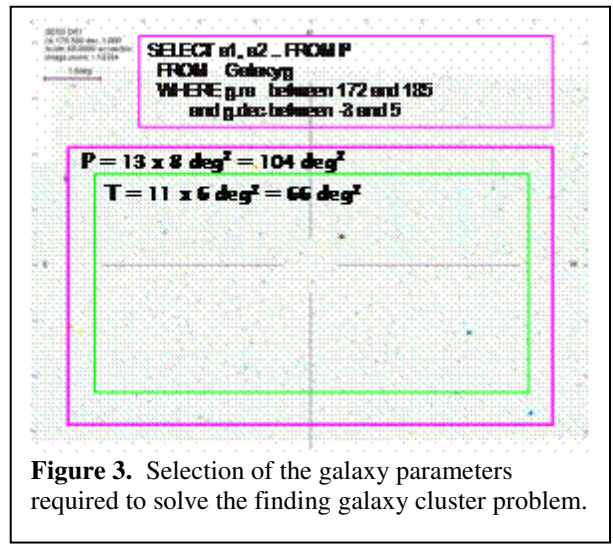

**Figure 3.** Selection of the galaxy parameters required to solve the finding galaxy cluster problem.

The SQL application does not extract the data to files prior to doing the processing. It uses the power of the database system to SELECT the necessary data and to do some processing and filtering inside the database. The processing requires basically one SELECT statement to extract the 5 parameters of interest from the general Galaxy table. Each of these rows or galaxies is JOINED with the 1000-row redshift lookup k-correction table to compute the BCG likelihood. This process eliminates candidates below some threshold very early in the computation.

These two steps are fairly simple and fast. The next step, counting the number of neighbors to estimate the BCG likelihood, is a bit more complex.

Neighborhood searches are usually very expensive because they imply computing distances between all pairs of objects in order to select those within some radius. Relational databases are well suited to look for objects meeting some criteria. However, when the searches are spatial, they usually require a special indexing system. We used the techniques described in [11] to perform the neighborhood searches. We tried both the Hierarchical Triangular Mesh (HTM) [12] and the zone-based neighbor techiniques. As explained below, the *Zone* index was chosen to perform the neighbor counts because it offered better performance.

The concept behind the zone-indexing schema is to map the celestial sphere into stripes of certain height called *Zones*. Each object at position (ra, dec) is assigned to a *Zone* by using the fairly simple formula $Zone = floor((dec + 90) / h)$, where $h$ is the *Zone* height.

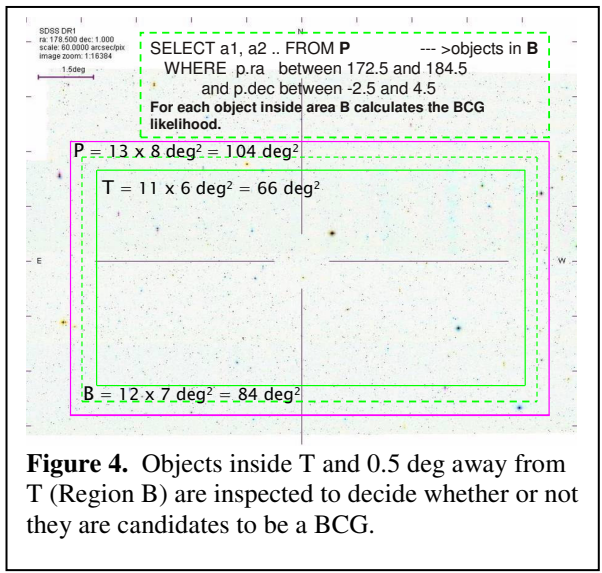

**Figure 4.** Objects inside T and 0.5 deg away from T (Region B) are inspected to decide whether or not they are candidates to be a BCG.

Zone-indexing has two benefits. First, using relational algebra the algorithm performs the neighborhood searches by joining a *Zone* with itself and discarding those objects beyond some radius. This pure SQL approach avoids the cost of using expensive calls to the external C-HTM libraries to do the spatial searches. Second, the data and computation partition very easily by assigning different *Zones* to each SQL Server and running the MaxBCG code in parallel.

The SQL MaxBCG algorithm works as follows. Given a target area T, all objects inside T and up to 0.5 deg away from T (buffer area B) are inspected to decide whether they are candidates to be the brightest cluster galaxy [Figure 4]. Searches for neighbors include all objects inside P which guarantees 0.5 deg buffer for objects near the border. This computation is therefore more accurate than the TAM version which used only a 0.25 deg buffer only. Area T differs from area B because deciding whether a candidate is the brightest cluster galaxy requires knowledge about candidate neighbors within 0.5 deg. To avoid unnecessary dependencies, we do in advance what will be required later. This task generates a **Candidates** table **C**.

In the next stage, all **candidate** galaxies in target area T are inspected to decide whether or not they have the maximum likelihood to be the brightest galaxy of their cluster. This neighbor search is done only among objects in the **Candidate** table, **C** [Figure 5]. This step creates a **Cluster** catalog where the likelihood of all candidates has been properly computed using 0.5 deg buffer around each candidate.

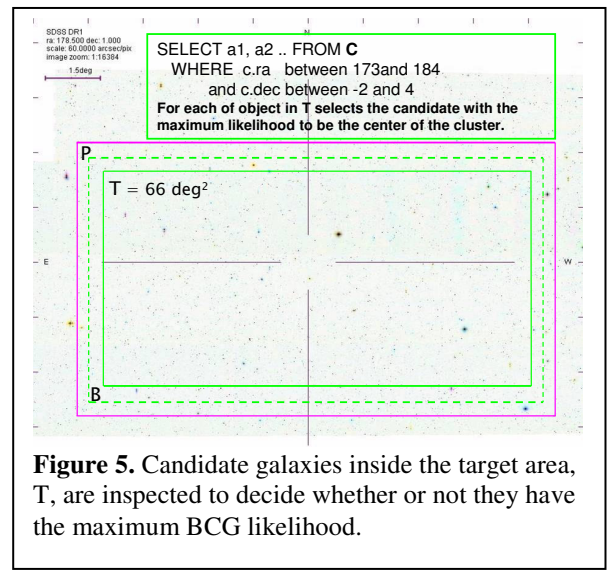

**Figure 5.** Candidate galaxies inside the target area, T, are inspected to decide whether or not they have the maximum BCG likelihood.

Processing a target field of 66 deg$^2$ as described above, requires about 5 hours with a dual 2.6 GHz machine running Microsoft SQL Server 2000. However, SQL Server is usually I/O bound instead of CPU bound so algorithm performance will not scale exactly with CPU speed.

| Table 1. SQL Server cluster performance, with no partitioning and with 3-way partitioning. | | | | | |
|---|---|---|---|---|---|
|  | Task | elapse (s) | cpu (s) | I/O | Galaxies on each partition |
| **No Partitioning** | spZone | 563.7 | 210.2 | 102,144 |  |
|  | fBCGCandidate | 15,758.2 | 15,161.0 | 562 |  |
|  | fIsCluster | 2,312.7 | 6,58.5 | 16,043 |  |
|  | total | **18,635** | **16,030** | **118,749** | **1,574,656** |
| **3-node Partitioning** | | | | | |
| P1 | spZone | 285.5 | 65.5 | 46,758 |  |
|  | fBCGCandidate | 6,099.1 | 5,850.7 | 209 |  |
|  | fIsCluster | 286.6 | 189.4 | 2,910 |  |
|  | total | **6,671.2** | **6,105.6** | **49,877** | **729,234** |
| P2 | spZone | 325.4 | 77.9 | 50,519 |  |
|  | fBCGCandidate | 8,210.7 | 7,907.7 | 306 |  |
|  | fIsCluster | 451.8 | 306 | 476 |  |
|  | total | **8,987.9** | **8,291.6** | **51,301** | **898,916** |
| P3 | spZone | 326.3 | 65.6 | 46,275 |  |
|  | fBCGCandidate | 6,121.5 | 5,783.5 | 283 |  |
|  | fIsCluster | 189.4 | 158.1 | 1,955 |  |
|  | total | **6,637.2** | **6,007.2** | **48,513** | **719,900** |
| **Partitioning Total** |  | **8,988** | **20,404** | **149,691** | **2,348,050** |
| **Ratio 1node/3node** |  | **48%** | **127%** | **126%** |  |

Resolving the same target area of 66 deg$^2$ with only one of the TAM CPUs using the file-oriented approach required about 73 hours (1000 s per each 0.25 deg$^2$ field), but that computation had only a 0.25 deg surrounding buffer and only 100 redshift steps. TAM would require about 25 times longer to do the equivalent SQL-calculation with a 0.5 deg buffer and redshift steps of 0.001.

### 2.4 SQL Server Cluster

The SQL implementation can run either on a single SQL Server or on a cluster of SQL Servers. As mentioned before, the problem is intrinsically parallel; each target area T can be processed in parallel. Using the *Zone* strategy described in section 2.3, a single target area may be processed in parallel by distributing the *Zones* among several servers allowing parallel execution of MaxBCG on different partitions of the target area [Figure 6].

When running in parallel, the data distribution is arranged so *each server is completely independent* from the others. We achieve this by duplicating some data and processing on different servers. The duplicated computations are insignificant compared to the total work involved when processing big volumes of data, or equivalently, big areas of the sky. We benchmarked this partitioning approach using a Microsoft SQL Server 2000 cluster composed of 3 nodes, each one a dual 2.6 GHz Xeon with 2 GB of RAM.

Table 1 shows the elapsed times, CPU times, and I/O operations used by SQL Server when solving MaxBCG with and without partitioning. **SpZone** is the task that arranges the data in *Zones* so the neighborhood searches are efficient. This task assigns a ZoneID and creates a clustered-index on the data. **fBCGCandidate** is the main task. It includes the BCG likelihood computations. Here is where the main neighborhood searches are performed to estimate properly the BCG likelihood. The fact that the I/O density is low during **fBCGCandidate** indicates the required data is usually in memory, which is always highly desired. Finally, **fIsCluster** screens the **Candidates** table and decides whether or not a candidate is a BCG. Although not included in Table 1, we also have the function that collects the galaxies that belong to a cluster. This is a fairly simple and fast operation which searches for neighboring galaxies within some radius for each detected cluster.

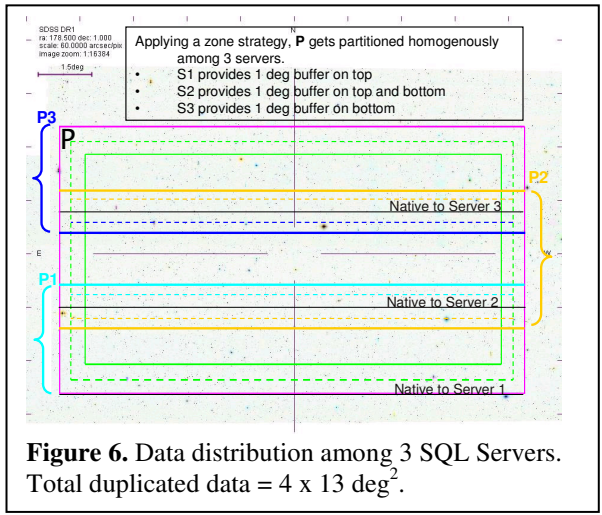

**Figure 6.** Data distribution among 3 SQL Servers. Total duplicated data = 4 x 13 deg$^2$.

The union of the answers from the three partitions is identical to the BCG candidates and clusters returned by the sequential (one node) implementation. Overall the parallel implementation gives a 2x speedup at the cost of 25% more CPU and I/O (including the cost of rezoning).

### 2.5 Time Performance

Tables 2 and 3 present a side-by-side comparison showing that the relational database solution is about 40 times faster per node than the file-based approach. For the specific cluster configurations considered here the 3-node SQL Server approach is about 20 times faster than the 5-node TAM.

Even if one were willing to wait 20 times longer, TAM nodes do not have enough memory to handle z-steps of 0.001 and a buffer of 0.5 deg. As mentioned before, a single TAM CPU takes 1000 s to process a target field of 0.25 $deg^2$ with a buffer of 0.25 deg and z-steps of 0.01 TAM performance is expected to scale lineally with the number of fields.

**Table 2.** Time scale factors for converting the TAM test case to the SQL server test case.

|  | TAM | SQL Server | Scale Factor |
|---|---|---|---|
| CPUs used | 1 | 2 | 0.5 |
| CPU | 600 MHz | 2.6 GHz | ~ 0.25 |
| Target field | 0.25 $deg^2$ | 66 $deg^2$ | 264 |
| z- steps | 0.01 | 0.001 | 25 |
| Buffer | 0.25 deg | 0.5 deg | |
| **Total Scale Factor** | | | **825** |

Table 2 compares both configurations and provides the scale factor to convert the TAM test case into the SQL test case. We normalize for the fact that the TAM CPU is about 4 times slower by dividing by 4 -- in fact much of the time is spent waiting for disk so this is being generous to the TAM system which had a comparable disk subsystem. Even with that the ratio is about 2 hours to about 2 days.

**Table 3.** Scaled TAM vs. Measured SQL Server performance for a target field of 66 $deg^2$.

| Cluster | Nodes | Time(s) | Ratio |
|---|---|---|---|
| TAM | 1 | 825,000 | 44 |
| SQL Server | 1 | 18,635 | |
| TAM | 5 | 165,000 | 18 |
| SQL Server | 3 | 8,988 | |

### 2.6 Performance Analysis

*What makes things run faster in SQL than in the file-based application?* We wish we knew but we can no longer run the original code so we can only make educated guesses (one of the authors wrote the original code).

First, the SQL implementation discards candidates early in the process by doing a natural JOIN with the k-correction table and filtering out those rows where the likelihood is below some threshold. This reduces the number of operations for subsequent INNER JOINs with the k-correction table and other tables. The SQL design uses the redshift index as the JOIN attribute which speeds the execution. So, early filtering and indexing are a big part of the answer. Second, the main advantage comes from using the *Zone* [11] strategy to index the data and speed up the neighborhood searches.

The SQL design could be further optimized. The iteration through the galaxy table uses SQL cursors which are very slow. But there was no easy way to avoid them. Our tests used a galaxy table of roughly 1.5 million rows (44 bytes each). About 1.2 million of those galaxies need to be joined with the k-correction table (1000 rows x 40 bytes). Joining this in memory would require at least 80 GB. A possible optimization is to define some sort of sky partitioning algorithm that breaks the sky in areas that can fit in memory, 2 GB in our case. Once an area has been defined, the MaxBCG task is scheduled for execution. This approach would be similar to the cluster implementation described in section 2.4 but at the level of cluster nodes since different computer may have different memory resources.

## 3. Discussion

This work demonstrates that using a relational database management system and SQL can improve computational performance on data-intensive applications. But performance is not the only advantage of using general database management systems rather than implementing custom applications. There is no magic in a relational DBMS; anything it does can also be done in a custom application (e.g. one implemented in TCL and C!). In fact, a quality custom solution should outperform a general-purpose DBMS.

The SQL implementation of MaxBCG was considerable simpler than the Tcl-Astrotools implementation primarily because it leveraged the features of the SQL system for data access, indexing, and parallelism.

The scientist, in our case an astronomer, should be free to focus on the science and minimize the effort required to optimize the application. Database management systems are designed to do fast searches, workload balancing and manage large data volumes and certainly will do a better job compared to what an average scientist could code. Database management systems allow simultaneous data access from different applications providing a good sharing environment.

So, the first lesson to learn for scientists working in data-intensive disciplines like astronomy, biology, etc. is that database systems are powerful tools to analyze big volumes of data and share results with others. On the other hand, the community researching database systems should ask itself why scientists are so reluctant to use database technologies.

As stated in the introduction, although the potential benefits of using database systems on the Grid has been recognized [16], their actual use as analysis tools is minimal. To our knowledge, most of the data-intensive applications that run on the Grid today focus on moving hundreds of thousands of files from the storage archives to the thousands of computing nodes. Many of these applications, like the one described in this paper, could solve the same problem more efficiently using databases.

We believe there is a basic reason for the absence of database technology in the Grid science community. While it is relatively easy to deploy and run applications coded in C, Fortran, Tcl, Python, Java, etc.; it is difficult to find resources to do the equivalent tasks using databases. Grid nodes hosting big databases and facilities where users can have their own database with full power to create tables, indexes, stored procedures, etc. are basically nonexistent. However, such facilities are needed to minimize the distance between the stored data and the analysis nodes, and in this way to guarantee that is the code that travels to the data and not the data to the code.

With the motivation of minimizing the distance between the SDSS CAS databases and analysis computing nodes, we implemented the SDSS Batch Query System, CasJobs [13][14]. The next section describes CasJobs and our work to develop an efficient Grid-enabled implementation of MaxBCG that instead of transferring hundreds of thousands of files over the network [6], leverages database technologies as parallel querying processing and indexing.

## 4. CasJobs, MaxBCG and Data Grids

CasJobs is an application available through the SkyServer site [15] that lets users submit long-running SQL queries on the CAS databases. The query output can be stored on the server-side in the user's personal relational database (MyDB). Users may upload and download data to and from their MyDB. They can correlate data inside MyDB or with the main database to do fast filtering and searches. CasJobs allows creating new tables, indexes, and stored procedures. CasJobs provides a collaborative environment where users can form groups and share data with others.

MaxBCG can be run using CasJobs, but that implementation is equivalent to the one described in section 2.3, which uses only one server. We want to take it one step further. Inspired by our SQL Server cluster experience, we plan to implement an application able to run in parallel using several systems. So for example when the user submits the MaxBCG application, upon authentication and authorization, the SQL code (about 500 lines) is deployed on the available Data-Grid nodes hosting the CAS database system. Each node will analyze a piece of the sky in parallel and store the results locally or, depending on the policy, transfer the final results back to the origin. We aim for a general implementation that makes it easy to bring the code to the data, avoids big data transfers, and extrapolates easily to solve other problems.

At the moment, two different organizations host the CAS database and the CasJobs system; Fermilab (Batavia, IL, USA) and The Johns Hopkins University (Baltimore, MD, USA). In the near future, the Inter-University Centre for Astronomy and Astrophysics (IUCCA) in Pune, India, will also host the system. Other organizations have showed interest in DB2 implementations of the CAS database. These are institutions with different access policies, autonomous and geographically distributed. CasJobs is accessible not only through the Web interface but also through Web services. Once the GGF DAIS protocol [21] becomes a final recommendation, it should be fairly easy to expose CasJobs Web services wrapped into the official Grid specification. We are working on issues of security, workflow tracking, and workload coordination, which need to be resolved to guarantee quality of service. Autonomy, geographical distribution, use of standards and quality of service are the key characteristics that a system requires in order to be accepted as a Grid system [27].

## 5. Conclusion

This paper presents a typical astronomical data-intensive application which aims to find galaxy clusters in SDSS catalog data. It demonstrates that using a database cluster achieves better performance than a file-based Tcl-C implementation run on a traditional Grid system. It also describes future work to "gridify" the implementation.

It points out that even though database systems are great tools for data-intensive applications, and even though one of the main goals of the Grid is providing infrastructure and resources for such applications, the are virtually no database management systems on the Grid to do effective data analysis.

In current Grid projects, databases and database systems are typically used only to access and integrate data, but not to perform analytic or computational tasks. Limiting usage in this manner neglects a strength of database systems, which is their ability to efficiently index, search, and join large amounts of data – often in parallel. It is a mistake to move large amounts of data to the query, when you can move the query to the data and execute the query in parallel. For this reason, it would be useful for nodes on the Grid to support different Database Management Systems so that SQL applications could be deployed as easily as traditional Grid applications coded in C, Fortran, etc.

## 6. Acknowledgments

This work is funded by the NASA Applied Information Systems Research Program. Grant NRA0101AISR. The paper benefited from comments by David DeWitt.

## References

[1] J. Annis, S. Kent, F. Castander, D. Eisenstein, J. Gunn, R. Kim, R. Lupton, R. Nichol, M. Postman, W. Voges, The SDSS Collaboration, "The maxBCG technique for finding galaxy clusters in SDSS data", *Bulletin of the American Astronomical Society, Vol. 31, p.1391*. December 1999.

[2] C. Stoughton, R. Lupton et al., "Sloan Digital Sky Survey: Early Data Release", *The Astronomical Journal, Volume 123, Issue 1, pp 485-548.* January 2002.

[3] G. Sergey, E. Berman, C. H. Huang, S. Kent, H. Newberg, T. Nicinski, D. Petravick, C. Stoughton, R. Lupton, "Shiva: an astronomical data analysis framework", *ASP Conf. Ser. 101: Astronomical Data Analysis Software and Systems V.* 1996.

[4] The Terabyte Analysis Machine Project Publications. http://tamhost.fnal.gov/Publications.html

[5] Condor: http://www.cs.wisc.edu/condor/

[6] J. Annis, Y. Zhao, J. Voeckler, M. Wilde, S. Kent, and I. Foster, "Applying Chimera Virtual Data Concepts to Cluster Finding in the Sloan Sky Survey", *Proceedings of the 2002 ACM/IEEE conference on Supercomputing,* Baltimore, MD, USA, November 16 - 22, 2002, pp 1 – 14.

[7] International Virtual Observatory Alliance. http://www.ivoa.net

[8] A. Szalay, J. Gray, "The World-Wide Telescope". *Science*, *293*, 2037-2040.

[9] Gladders, Michael D., Yee, H. K. C, "A New Method For Galaxy Cluster Detection. I. The Algorithm", *The Astronomical Journal, Volume 120, Issue 4, pp. 2148-2162*. October, 2000.

[10] J. Gray, A.S. Szalay, A. Thakar, P. Kunszt, C. Stoughton, D. Slutz, J. vandenBerg, "The Sloan Digital Sky Survey Science Archive: Migrating a Multi-Terabyte Astronomical Archive from Object to Relational DBMS", *Distributed Data & Structures 4: Records of the 4th International Meeting*, pp 189-210 W. Litwin, G. Levy (eds), Paris France March 2002, Carleton Scientific 2003, ISBN 1-894145-13-5, also *MSR-TR-2002-01*, January 2002.

[11] J. Gray, A. S. Szalay, A. R. Thakar, G. Fekete, W. O'Mullane, G. Heber, A. H. Rots, "There Goes the Neighborhood: Relational Algebra for Spatial Data Search", *Microsoft Technical Report MSR-TR-2004-32*. April, 2004. ftp://ftp.research.microsoft.com/pub/tr/TR-2004-32.pdf

[12] P. Z. Kunszt, A. S. Szalay, I. Csabai, A. R. Thakar, "The Indexing of the SDSS Science Archive", *ASP Conf. Ser., Vol 216, Astronomical Data Analysis Software and Systems IX* eds. Nadine Manset, Christian Veillet, and Dennis Crabtree (San Francisco: ASP), 2000, pp 141- 145.

[13] W. O'Mullane, J. Gray, N. Li, T. Budavari, M. Nieto-Santisteban, A. Szalay, "Batch Query System with Interactive Local Storage for SDSS and the VO", *ASP Conf. Ser., Vol 314, Astronomical Data Analysis Software and Systems XIII.* eds. Francois Ochsenbein, Mark G. Allen and Daniel Egret (San Francisco: ASP), 2004, pp 372-375. http://adass.org/adass/proceedings/adass03/O4-4/

[14] CasJobs http://casjobs.sdss.org/CasJobs/

[15] SkyServer http://skyserver.sdss.org

[16] P. Watson. "Databases and the Grid", *UK e-Science Programme Technical Report UKeS-2002-01, Natioanl e-Science Centre.* http://www.nesc.ac.uk/technical_papers/PaulWatsonDatabasesAndTheGrid.pdf

[17] M. L. Bote-Lorenzo, Y. A. Dimitriadis, and E. Gómez-Sánchez. "Grid Characteristics and Uses: A Grid Definition", *Across Grids 2003, LNCS 2970, pp 291-298, 2004*. eds F. Fernández Rivera et al. (Springer-Verlag Berlin Heidelberg 2004)

[18] I. Foster and C. Kesselman. *The Grid: Blueprint for a NewComputing Infrastructure*. Morgan Kaufmann Publishers, San Francisco, CA, USA, second edition, 2003.

[19] D. Pearson, "Data Requirements for the Grid", Technical Report, 2002. http://www.cs.man.ac.uk/grd-db/papers/Requirements.pdf

[20] N. Paton, M. Atkinson, V. Dialani, D. Pearson, T. Storey, and P. Watson, "Databases Access and Integration Services on the Grid." *UK e-Science Programme Technical Report UKeS-2002-03. National e-Science Centre.* http://www.nesc.ac.uk/technical_papers/dbtf.pdf

[21] DAIS-WG https://forge.gridforum.org/projects/dais-wg

[22] OGSA-DAI http://www.ogsa-dai.org.uk/

[23] OGSA-DQP http://www.ogsa-dai.org.uk/dqp/

[24] The Chimera Virtual Data System. http://www.griphyn.org/chimera/

[25] Pegasus. http://pegasus.isi.edu/

[26] The Metadata Catalog Service. http://www.isi.edu/~deelman/MCS/

[27] I. Foster , "What is the Grid? A Three Point Checklist" *Grid Today July 22, 2002: VOL. 1 NO. 6* http://www.gridtoday.com/02/0722/100136.html

[28] J. Binney, M. Merrifield , *Galactic Astronomy*, Princeton University Press, ISBN: 0691004021 1998.

[29] MaxBCG. http://skyservice.pha.jhu.edu/develop/applications/MaxBCG.aspx

## MaxBCG SQL code for MySkyServerDr1 (http://www.skyserver.org/myskyserver/)
## Date: Nov / 23 / 2004

**Note:** If you wish to try this code using CasJobs (http://casjobs.sdss.org/casjobs),
substitute MySkyServerDr1.dbo for your target database (e.g: dr1, dr2, or dr3)
If you have MyDB or Interface problems please contact
  Nolan Li <nli@pha.jhu.edu>,
  Wil O'Mullane <womullan@skysrv.pha.jhu.edu>

**General questions** about the SQL code to Maria A. Nieto-Santisteban nieto@pha.jhu.edu

```sql
-- ******************************* Schema
CREATE TABLE Kcorr (          --/D expected brightness and color of a BCG at given redshift
       zid    int identity (1,1) PRIMARY KEY NOT NULL,
       z      real,           --/D redshift
       i      real,           --/D apparent i petro mag of the BCG @z
       ilim   real,           --/D limiting i magnitude @z
       ug     real,           --/D K(u-g)
       gr     real,           --/D K(g-r)
       ri     real,           --/D K(r-i)
       iz     real,           --/D K(i-z)
       radius float            --/D radius of 1Mpc @z
)
-- Import the K-correction table into your database

CREATE TABLE Galaxy (         --/D One row per SDSS Galaxy, extracted from PhotoObjAll
       objid   bigint PRIMARY KEY,   --/D Unique identifier of SDSS object
       ra      float,          --/D Right ascension in degrees
       dec     float,          --/D Declination in degrees
       i       real,           --/D Magnitude in i-band
       gr      real,           --/D color dimension g-r
       ri      real,           --/D color dimension r-i
       sigmagr float,          --/D Standard error of g-r
       sigmari float            --/D Standard error of r-i
)

CREATE TABLE Candidates (     --/D The list of BCG candidates
       objid   bigint PRIMARY KEY,   --/D Unique identifier of SDSS object
       ra      float,          --/D Right ascension in degrees
       dec     float,          --/D Declination in degrees
       z       float,          --/D redshift
       i       real,           --/D magnitude in the i-band
       ngal    int,            --/D number of galaxies in the cluster
       chi2    float            --/D chi-squared confidence in cluster
)

CREATE TABLE Clusters (       --/D Selected BCGs from the candidate list
       objid   bigint PRIMARY KEY,   --/D Unique identifier of SDSS object
       ra      float,          --/D Right ascension in degrees
       dec     float,          --/D Declination in degrees
       z       float,          --/D redshift
       i       real,           --/D magnitude in the i band
       ngal    int,            --/D number of galaxies in the cluster
       chi2    float            --/D chi-squared confidence in cluster
)

CREATE TABLE ClusterGalaxiesMetric (--/D Cluster galaxies inside 1 MPc at R200
       clusterObjID   bigint,--/D BCG unique identifier (cluster center)
       galaxyObjID    bigint,--/D Galaxy unique identifier (galaxy part of the cluster)
       distance       float  --/D distance between cluster and galaxy
)
GO

CREATE VIEW Zone AS    --/D Primary Galaxy view of the zone table in SDSS database.
SELECT  ZoneID,               --/D Zone number based on 30 arcseconds
        objid,                --/D Unique identifier of SDSS object
        ra,                   --/D Right ascension in degrees
        dec,                  --/D Declination in degrees
        cx,                   --/D x, y, z unit vector of object on celestial sphere
        cy,                   --/D
        cz                    --/D
FROM MySkyServerDr1.dbo.Zone   --/D
WHERE mode = 1 and type = 3    --/D Primary and Galaxy
-- ******************************* End Schema
GO
```

```sql
-------------------------------------------------
CREATE FUNCTION fGetNearbyObjEqZd(@ra float, @dec float, @r float)
-------------------------------------------------
--/H Returns a table of objects from the Zone view (here Primary Galaxies)
--/H within @r degrees of an Equatorial point (@ra, @dec)
--/A
-------------------------------------------------
--/T Table has format (objID bigint, distance float (degrees))
--/T <samp>
--/T <br> select * from fGetNearbyObjEqZd(2.5, 3.0,0.5)
--/T </samp>
-------------------------------------------------
RETURNS @neighbors TABLE (ObjID bigint, distance float) AS
BEGIN
  DECLARE
        @zoneHeight     float,  --/D standard scale height of SDSS zone
        @zoneID         int,    --/D loop counter
        @cenZoneID      int,    --/D Zone where the input (@ra, @dec) belongs (central zone)
        @maxZoneID      int,    --/D Maximum zone
        @minZoneID      int,    --/D Minimum zone
        @adjustedRadius real,   --/D Radius adjusted by cos(dec)
        @epsilon        real,   --/D Small value to avoid division by zero
        @r2             float,  --/D squared radius
        @x              float,  --/D used in ra cut to minimize searches in upper and lower
                                --/D zones within the search radius
        @dec_atZone     float,  --/D max dec for Zones below central zone
                                --/D min dec for Zones above the central zone
        @delta_dec      float,  --/D distance between declination and dec_atZone,
                                --/D necessary to compute @x
        @zoneID_x       int,    --/D zoneID to compute @x
        @cx             float,  --/D Input's Cartesian coordinates
        @cy             float,
        @cz             float,
        @d2r            float;  --/D PI()/180.0, from degrees to radians
  SET @zoneHeight = 30.0 / 3600.0; -- 30 arcsec in degrees
  SET @d2r = PI()/180.0          -- radian conversion
  SET @epsilon = 1e-9            -- prevents divide by zero
  SET @cx = COS(@dec * @d2r) * COS(@ra * @d2r) -- convert ra,dec to unit vector
  SET @cy = COS(@dec * @d2r) * SIN(@ra * @d2r)
  SET @cz = SIN(@dec * @d2r)  -- radial distance measured in degrees is larger away from the equator
  SET @adjustedRadius = @r / (COS(RADIANS(ABS(@dec))) + @epsilon) -- adjustRadius corrects for this.
  SET @r2 = 4 * POWER(SIN(RADIANS(@r/2)),2) -- Assumes input radius in degrees
  -------------------------------------------------
  -- loop over all zones that overlap the circle of interest looking for objects inside circle.
  SET @cenZoneID = FLOOR((@dec + 90.0)      / @zoneHeight) -- zone holding ra,dec point
  SET @maxZoneID = FLOOR((@dec + @r + 90.0) / @zoneHeight) -- max zone to examine
  SET @minZoneID = FLOOR((@dec - @r + 90.0) / @zoneHeight) -- min zone to examine
  SET @zoneID = @minZoneID
  WHILE (@zoneID <= @maxZoneID)        -- Loop through all zones from the bottom to the top
    BEGIN
        IF (@zoneID = @cenZoneID)      -- first compute @x which further restricts the ra range
          SET @x = @adjustedRadius     -- within a zone.  The circle is narrower in
        ELSE                           -- zones away from the center zone, and x gives this
          BEGIN                        -- narrowing factor (measured in degrees)
             SET @zoneID_x = @zoneID
             IF (@zoneID < @cenZoneID)
               SET @zoneID_x = @zoneID_x + 1
             SET @dec_atZone = @zoneID_x * @zoneHeight - 90 -- Zones below the center zone will get
                                                           -- the max dec in the zone, Zones above will get
                                                           -- the ~min dec in the zone
             SET @delta_dec = ABS(@dec - @dec_atZone) -- how far away is the zone border?
             SET @x =  SQRT(ABS(POWER(@r,2)-
                            POWER(@delta_dec,2))) /
                     (COS(RADIANS(ABS(@dec_atZone))) + @epsilon) -- adjust @x for declinations away
                                                                 -- from the equator
          END
        INSERT @neighbors              -- now add in the objects of this zone that are inside circled
          SELECT  objID,               -- the id of the nearby galaxy
                 SQRT(POWER(cx - @cx, 2) +
                      POWER(cy - @cy, 2) +
                      POWER(cz - @cz, 2)
                      ) / @d2r AS distance -- in degrees
          FROM ZONE                    -- ZONE View of primary galaxies
          WHERE zoneID = @zoneID       -- using zone number and ra interval
            AND ra  BETWEEN @ra - @x AND @ra + @x
            AND dec BETWEEN dec - @r AND dec + @r
            AND @r2 > POWER(cx - @cx, 2) + POWER(cy - @cy, 2) + POWER(cz - @cz, 2)
        SET @zoneID = @zoneID +1      -- next zone
    END                               -- bottom of the loop
  RETURN
END     -- ******************************* fGetNearbyObjEqZd
GO
```

```sql
----------------------------------------------------
CREATE FUNCTION fBCGCandidate( --D Calculates the BCG likelihood
        @objid bigint,        --D Unique identifier of SDSS object
        @ra    float,         --D Right ascension in degrees
        @dec   float,         --D Declination in degrees
        @imag  real)          --D i-band magnitude
----------------------------------------
--/H Returns a table of BCG candidate likelihoods of neighbors of a given object
--/A
----------------------------------------------------------
------------------------------------------------------------
--/H If the input galaxy is likely to be a BCG at any resdshift
--/H this function returns the position, redshift, number of galaxies,
--/H and best chisquare estimation.
--/H The table returned may have zero or one rows
----------------------------------------------------
RETURNS @t TABLE (
             objid  bigint, --D Unique identifier of SDSS object
             ra     float,  --D Right ascension in degrees
             dec    float,  --D Declination in degrees
             z      float,  --D estimated redshift from the K-correction
             ngal   int,    --D number of galaxies in the neighborhood
             chi2   float   --D chi square estimate
)
AS
BEGIN
  DECLARE
       @rad    float,       --D Search radius
       @imin   real,        --D minimum magnitude in the i-band
       @imax   real,        --D maximum magnitude in the i-band
       @grmin  real,        --D minimum g-r color magnitude
       @grmax  real,        --D maximum g-r color magnitude
       @rimin  real,        --D minimum r-i color magnitude
       @rimax  real,        --D maximum r-i color magnitude
       @chi    float,       --D minimum estimated chi square error
       @grPopSigma real,    --D g-r constant to estimate chi square
       @riPopSigma real     --D r-i constant to estimate chi square

  SET @grPopSigma = 0.05;
  SET @riPopSigma = 0.06

  DECLARE @chisquare TABLE ( --D This temporary table contains an object, at all redshifts,
                             --D where is likely to be a BCG (may have more than one row)
                             --D It is the result of JOIN with the k_correction table and
                             --D further filtering
             zid    int PRIMARY KEY NOT NULL,
             z      real,  --D redshift
             i      real,  --D i-band magnitude
             chisq  float, --D chisq estimate
             ngal   int    --D number of galaxies
  )

  DECLARE @friends TABLE (   --D Neighbors of the object being processed
             objid   bigint, --D Unique identifier of SDSS object
             distance float, --D Distance in degrees
             I       real,   --D i-band magnitude
             gr      real,   --D g-r color
             ri      real    --D r-i color
  )

  DECLARE @counts TABLE (            --D Keeps record of number of galaxies per redshift
             zid    int PRIMARY KEY NOT NULL,   --D redshift ID
             ngal   int                         --D Number of galaxies
  )
```

```sql
    -- body of fBCGCandidate() function
    --==================================
    -- Filter step: Calculates the unweighted BCG likelihood and discards unlikely BCGs
    INSERT @chisquare
      SELECT k.zid,              -- the redshift ID
             k.z,                -- the flux in z and i bands
             g.i,                -- and the chi squared estimator
             POWER(g.i-k.i,2) / POWER (0.57,2)   +
             POWER (g.gr - k.gr,2) / (POWER (sigmagr,2) + POWER (@grPopSigma,2)) +
             POWER (g.ri - k.ri,2) / (POWER (sigmari,2) + POWER (@riPopSigma,2)) AS chisq,
             0 AS ngal
      FROM Galaxy g CROSS JOIN Kcorr k
      WHERE objid = @objid
        AND(POWER (g.i-k.i,2) / POWER (0.57,2) + -- 0.57 is the population dispersion of BCG magnitudes
            POWER (g.gr - k.gr,2) / (POWER (sigmagr,2) + POWER (@grPopSigma,2)) +
            POWER (g.ri - k.ri,2) / (POWER (sigmari,2) + POWER (@riPopSigma,2))
           ) < 7
    --================================
    -- If the galaxy passed the filter at some redshift, then evaluate it.
    IF @@rowcount > 0
      BEGIN
            -- Calculate window values for magnitudes and colors from the k-correction table
            SELECT @imin=@imag;
            SELECT  @rad    = MAX (k.radius),     -- the maximum angular radius of 1 Mpc
                    @chi    = MIN (chisq),        -- the chi squared estimator
                    @imax   = MAX (k.ilim),       -- add correct shift
                    @grmin  = MIN (k.gr) - 2*@grPopSigma,
                    @grmax  = MAX (k.gr) + 2*@grPopSigma,
                    @rimin  = MIN (k.ri) - 2*@riPopSigma,
                    @rimax  = MAX (k.ri) + 2*@riPopSigma
            FROM @chisquare c JOIN Kcorr k ON c.zid = k.zid
            -- Look for neighbors in the Zone table with similar magnitudes and colors.
            -- Retrieves other attributes by joining with Galaxy
            INSERT @friends
              SELECT n.objid, n.distance, g.i, g.gr, g.ri
              FROM fGetNearbyObjEqZd(@ra,@dec,@rad) n JOIN Galaxy g ON g.objid = n.objid
              WHERE n.objid != @objid
                AND g.i  BETWEEN @imin  AND @imax
                AND g.gr BETWEEN @grmin AND @grmax
                AND g.ri BETWEEN @rimin AND @rimax
            -- Count the number of galaxies with similar magnitudes and colors grouped by redshfit
            INSERT @counts
              SELECT c.zid, COUNT(*) AS ngal
              FROM @chisquare c  JOIN Kcorr k ON c.zid = k.zid
                   CROSS JOIN  @friends f
              WHERE f.distance < k.radius
                AND f.i  BETWEEN @imag AND k.ilim
                AND f.gr BETWEEN k.gr - @grPopSigma AND k.gr + @grPopSigma
                AND f.ri BETWEEN k.ri - @riPopSigma AND k.ri + @riPopSigma
              GROUP BY c.zid
            -- Update the counts in the chisquare table
            UPDATE @chisquare
              SET ngal= q.ngal
              FROM @chisquare c, @counts q
              WHERE c.zid = q.zid

            -- Weight the chisquare and select the maximum
            -- It must have at least one neighbor
            SELECT @chi = MAX (LOG(ngal+1) - chisq)
            FROM @chisquare
            WHERE ngal>0
            -- Return estimated redshift, number of neighbors and likelihood
            IF @chi IS NOT NULL
              BEGIN
                    INSERT @t
                      SELECT
                            @objid AS objid, @ra AS ra,  @dec AS dec,
                            z,                    -- redshift
                            ngal+1 AS ngal,       -- number of neighbors
                            @chi AS chi2          -- likelihood
                      FROM @chisquare
                      WHERE ABS (LOG(ngal+1) - chisq - @chi) < 0.00000001
              END
      END
    RETURN
END
-- ******************************** fGetCandidate
GO
```

```sql
-----------------------------------------
CREATE FUNCTION fBCGr200(@ngal float)
-----------------------------------------
--/H Returns the r200 radius in Mpc.
--/H The mean density inside the r200 radius is 200 times the mean galaxy density of the sky
RETURNS float
AS
BEGIN
        RETURN 0.17 * POWER(@ngal,0.51);
END
--********************************* fBCGr200
GO
-------------------------------------------------
CREATE FUNCTION fIsCluster(@objid bigint,
        @ra float, @dec float, @z real, @ngal int, @chi2 float)
-------------------------------------------------
--/H returns 1 if this is a cluster center, 0 else
-------------------------------------------------
RETURNS int
AS
BEGIN
  DECLARE @rad float, @chi float;
  -- the r200 radius is, at ngal=100, 1.78 degree which, at z=0.05, is 0.74 degrees.
  -- So, the maximum size needed for chiSq (BCG) calculations is 0.75 degrees
  -- from the edge of the region to be coalesced.
  SELECT @rad = radius
  FROM Kcorr
  WHERE ABS(z - @z) < 0.0000001
  -- Select the best chi2 from candidate neighbors
  SELECT @chi = MAX(c.chi2)
  FROM fGetNearbyObjEqZd(@ra, @dec,@rad) n
    JOIN Candidates c ON n.objid = c.objid
  WHERE  c.z BETWEEN @z - 0.05 AND @z + 0.05;
  -- If the best chi2 corresponds to the input object then it is selected as the center
  RETURN
        CASE WHEN abs(@chi - @chi2) < 0.00001 THEN 1  ELSE 0 END
END
--********************************* fIsCluster
GO
------------------------------------------------------
CREATE FUNCTION fGetClusterGalaxiesMetric(@objid bigint,
        @ra float, @dec float, @z real, @imag real, @ngal float)
------------------------------------------------------
RETURNS @t TABLE ( clusterObjID bigint, galaxyObjID bigint, distance float
)
AS
BEGIN
  DECLARE
        @rad     float,
        @gr      real,
        @ri      real,
        @ilim    real,
        @grPopSigma real,
        @riPopSigma real
  SET @grPopSigma = 0.05;
  SET @riPopSigma = 0.06;
  --
  SELECT       @rad = radius * dbo.fBCGr200(@ngal),
               @ilim = ilim,
               @gr = gr, @ri=ri
  FROM Kcorr
  WHERE ABS (z - @z) < 0.0000001
  -- insert central galaxy first
  INSERT @t
    SELECT @objid AS clusterObjID, @objid AS galaxyObjID, 0 AS distance
  -- insert all the other "friends"
  INSERT @t
    SELECT @objid AS clusterObjID, n.objid AS galaxyObjID, n.distance
    FROM fGetNearbyObjEqZd(@ra,@dec,@rad) n
         JOIN Galaxy g  ON g.objid = n.objid
    WHERE n.objid != @objid
       AND n.distance < @rad
       AND g.i  BETWEEN @imag - 0.001 AND @ilim
       AND g.gr BETWEEN @gr - @grPopSigma AND @gr + @grPopSigma
       AND g.ri BETWEEN @ri - @riPopSigma AND @ri + @riPopSigma
  RETURN
END
--********************************* fGetClusterGalaxies
GO
-- ********************************* End Functions
```

```sql
-- ********************************** Stored Procedures
--------------------------------------
CREATE PROCEDURE spImportGalaxy (
--------------------------------------
--/H Import the data from the main Galaxy table into the MyDB Galaxy Table
--------------------------------------
        @minRa  float,
        @maxRa  float,
        @minDec float,
        @maxDec float
)
AS
BEGIN
  TRUNCATE TABLE Galaxy
  INSERT Galaxy
    SELECT objid,
           ra,
           dec,
           dered_i AS i,
           dered_g - dered_r AS gr,
           dered_r - dered_i AS ri,
           CAST(2.089 * POWER(10.000, 0.228 * dered_i-6.0) AS float) AS sigmagr,
           CAST(4.266 * POWER(10.0000,0.206 * dered_i-6.0) AS float) AS sigmari
      FROM MySkyServerDR1.dbo.Galaxy
     WHERE  ra  BETWEEN @minRA  AND @maxRa
       AND dec BETWEEN @minDEc  AND @maxDec
END
--********************************** spImportGalaxy
GO

--------------------------------------
CREATE PROCEDURE spMakeCandidates(
--------------------------------------
--/H Calls the fGetBCGCandidate function for each galaxy
--/H inside the limits @minRa, @maxRa, @minDec, @maxDec
--/H Fills Candidates table with BCG candidates
        @minRa  float,   -- the region of interest
        @maxRa  float,   -- ra and dec boundaries
        @minDec float,
        @maxDec float
)
AS
BEGIN
  SET NOCOUNT ON;
  TRUNCATE TABLE Candidates         -- empty the candidate table
  --
  DECLARE      @objid bigint,
               @ra float,
               @dec float,
               @imag real
  -- loop over all galaxies in the specified region, applying the fBCGCandidate() function to them
  DECLARE c CURSOR READ_ONLY
  FOR   SELECT g.objid, g.ra, g.dec, g.i
        FROM    Galaxy g
        WHERE   g.ra  BETWEEN @minRa  AND @maxRa
          AND   g.dec BETWEEN @minDec AND @maxDec
  OPEN c

  WHILE ( 1 = 1 )
    BEGIN
        FETCH NEXT FROM c INTO @objid, @ra, @dec, @imag
        IF (@@fetch_status < 0 ) BREAK
        INSERT Candidates
           SELECT objid, ra, dec, z,
                  @imag AS i,ngal, chi2
            FROM fBCGCandidate(@objid, @ra, @dec, @imag);
    END
  CLOSE c
  DEALLOCATE c
END
--********************************** spMakeCandidates
GO
```

```sql
-------------------------------------
CREATE PROCEDURE spMakeClusters
-------------------------------------
--/H Inserts BCG candidates into the Clusters table if they are the center of their cluster.
AS
BEGIN
  SET NOCOUNT ON;
  TRUNCATE TABLE Clusters                  -- empty the cluster table
  INSERT   Clusters                        -- insert candidates
    SELECT *
    FROM Candidates
    WHERE dbo.fIsCluster(objid, ra, dec, z, ngal, chi2) = 1
END
--********************************* spMakeClusters
GO

-------------------------------------
CREATE PROCEDURE spMakeGalaxiesMetric
-------------------------------------
-- /H Creates the ClusterGalaxiesMetric table with centers and cluster members.
AS
BEGIN
  TRUNCATE TABLE ClusterGalaxiesMetric
  DECLARE
        @objid bigint,
        @ra float,
        @dec float,
        @z real,
        @imag real,
        @ngal float

  ----------------------------------
  DECLARE c CURSOR
  FOR  SELECT objid, ra, dec, z, i, ngal
        FROM Clusters
  -- Loop over all clusters building the members from the center.
  OPEN c
  --
  WHILE ( 1 = 1 )
    BEGIN
        FETCH NEXT FROM c INTO @objid, @ra, @dec, @z, @imag, @ngal
        IF (@@fetch_status < 0 ) BREAK
        INSERT ClusterGalaxiesMetric
          SELECT *
          FROM fGetClusterGalaxiesMetric(@objid,@ra, @dec, @z, @imag, @ngal)
    END
    --
  CLOSE c
  DEALLOCATE c
END
-- ********************************* spMakeGalaxies
GO
-- ******************************* End Stored Procedures

--*********************************
-- MySkyServerDr1 covers about 2.5 x 2.5 deg^2 centered in 195.163 and 2.5
EXEC spImportGalaxy 190, 200, 0, 5 -- This will import the whole galaxy table
EXEC spMakeCandidates 194, 196, 1.5, 3.5

-- Our target 66 deg^2 inside 104 deg^2 buffer
-- EXEC spImportGalaxy 172, 185, -3, 5
-- EXEC spMakeCandidates 172.5, 184.5, -2.5, 4.5

EXEC spMakeClusters
EXEC spMakeGalaxiesMetric
--*********************************
```